%% file: main.tex
\documentclass[conference, final]{IEEEtran}
\IEEEoverridecommandlockouts
\usepackage{hyperref}       
\usepackage{url}            
\usepackage{booktabs}       
\usepackage{amsfonts}       
\usepackage{nicefrac}       
\usepackage{microtype}      
\usepackage{xcolor}         
\usepackage{multirow}
\usepackage{listings}
\usepackage{pgfplots}
\usepgfplotslibrary{dateplot}
\pgfplotsset{compat=1.17}
\usepackage{subcaption}
\usepackage{hyperref}
\usepackage{enumitem}
\usepackage{algorithm}
\usepackage{hyperref}
\usepackage{graphicx}
\usepackage{tikz-cd}
\usepackage{tcolorbox}
 \usepackage[frozencache,cachedir=mint]{minted}
 \usepackage{fancyvrb}
\usepackage{pgfplots}
\usepackage{xurl}
\usepackage[noadjust]{cite}
\usepackage{graphicx}
\usepackage{algpseudocode}
\usepackage{algorithm}
\usepackage{tabularx}
\usepackage{balance}
\usepackage{caption}
\usepackage{float}
\usepackage{makecell}
\usepackage{tikz}
\usepackage{pgfplots}
\pgfplotsset{compat=1.17}
\pgfdeclarelayer{background}
\pgfsetlayers{main}
\usepackage{multirow}
\usetikzlibrary{external}
\usepackage{rotating}
\usepackage{amsmath}

\usepackage{algorithm, algpseudocode}
\usepackage{setspace}
\usepackage{xcolor}
\usepackage{gensymb}
\usepackage{placeins}
\usepackage{enumitem}
\usetikzlibrary{decorations.pathreplacing,positioning}
\usepackage[noadjust]{cite}
\usepackage{dirtytalk}
\usepackage[table]{xcolor}
\usepackage{colortbl}

\usepgfplotslibrary{groupplots}

\newtcolorbox{devcomment}[2][]{%
  colback=gray!10,
  colframe=orange,
  fonttitle=\bfseries,
  title=#1:,
  #2
}

\pgfplotsset{compat=1.17}
 \usepackage{etoolbox}
\makeatletter
\patchcmd{\@verbatim}
  {\verbatim@font}
  {\verbatim@font\small}
  {}{}
\makeatother

\hypersetup{
   breaklinks=true,   
   colorlinks=true,   
}

\definecolor{archtBlue}{RGB}{0, 114, 178}
\definecolor{lightBlue}{RGB}{0, 114, 240}
\definecolor{archtRed}{RGB}{213, 94, 0}
\definecolor{lightRed}{RGB}{250, 94, 0}
\definecolor{archtGreen}{RGB}{0, 128, 0}
\definecolor{lightGreen}{RGB}{0, 230, 0}
\definecolor{archtPurple}{RGB}{128, 0, 128}
\definecolor{archtOrange}{RGB}{230, 159, 0}
\definecolor{archtPink}{RGB}{204, 121, 167}
\definecolor{customgreen}{RGB}{67, 102, 51}

\newcommand{\dataset}[0]{\textsc{ParaTrans}}

\newcommand{\hecbench}[0]{\textsc{HeCBench}}

\makeatletter
\newcommand*{\rom}[1]{\expandafter\@slowromancap\romannumeral #1@}
\makeatother

\def\BibTeX{{\rm B\kern-.05em{\sc i\kern-.025em b}\kern-.08em
    T\kern-.1667em\lower.7ex\hbox{E}\kern-.125emX}}

\DeclareRobustCommand*{\IEEEauthorrefmark}[1]{%
  \raisebox{0pt}[0pt][0pt]{\textsuperscript{\footnotesize\ensuremath{#1}}}}
  \usepackage[misc]{ifsym}

\ifCLASSINFOpdf

\else

\fi

\DeclareRobustCommand*{\IEEEauthorrefmark}[1]{%
  \raisebox{0pt}[0pt][0pt]{\textsuperscript{\footnotesize\ensuremath{#1}}}}
  \usepackage[misc]{ifsym}

\begin{document}

\title{\textit{UniPar}: A Unified LLM-Based Framework for Parallel and Accelerated Code Translation in HPC
}
\author{\IEEEauthorblockN{
\hspace{-1cm}Tomer Bitan\IEEEauthorrefmark{1},
Tal Kadosh\IEEEauthorrefmark{2,3},
Erel Kaplan\IEEEauthorrefmark{1},
Shira Meiri\IEEEauthorrefmark{1},
Le Chen\IEEEauthorrefmark{4},
Peter Morales\IEEEauthorrefmark{5},
Niranjan Hasabnis\IEEEauthorrefmark{5},
Gal Oren\IEEEauthorrefmark{1,6}}\\
\IEEEauthorblockA{\IEEEauthorrefmark{1}Technion, 
\IEEEauthorrefmark{2}Ben-Gurion University,
\IEEEauthorrefmark{3}IAEC,
\IEEEauthorrefmark{4}Argonne National Laboratory, 
\IEEEauthorrefmark{5}Code Metal,
\IEEEauthorrefmark{6}Stanford University}

{\tt\small tomerbitan@cs.technion.ac.il, talkad@post.bgu.ac.il, \{erel.kaplan,shira.meiri\}@campus.technion.ac.il,}\\
{\tt\small lechen@anl.gov, \{peter, niranjan\}@codemetal.ai, galoren@stanford.edu}\\
}
\makeatletter
\def\@IEEEpubidpullup{8\baselineskip}
\makeatother

\maketitle

\IEEEpubidadjcol

\IEEEpubidadjcol

\begin{abstract}


Translating programs between various parallel programming languages is an important problem in the high-performance computing (HPC) community, with implications for industry and academia. Existing tools for this problem are either too narrow in scope (translate between specific languages) and/or outdated (requiring maintenance). Recent explosive growth in the popularity of large language models (LLMs) and their ability to generate and translate code offers a potential alternative approach. Toward that end, we first need to systematically evaluate the ability of LLMs to translate between parallel languages.

In this work, we introduce \textbf{UniPar}, a systematic evaluation framework for LLM-based parallel code translation. Specifically, in this work, we target translations between serial code, CUDA, and OpenMP. Our goal is to assess how well current instruction-tuned LLMs -- specifically GPT-4o-mini and LLaMA-3.3-70B-Instruct -- can be used out of the box or enhanced through known strategies. We evaluated four major usage modes: hyperparameter optimization for decoding, zero- and few-shot prompting, supervised fine-tuning, and iterative feedback through compiler-based repair. As a part of the evaluation, we construct a new dataset called \dataset{}, covering both serial-to-parallel translation and cross-paradigm transformations.


Our findings reveal that while off-the-shelf models struggle under the default settings (e.g., GPT-4o-mini achieves only \textbf{46\%} compilation and \textbf{15\%} functional correctness), our UniPar methodology -- combining fine-tuning, hyperparameter tuning, and compiler-guided repair -- improves performance by up to 2X (\textbf{69\% compilation} and \textbf{33\% correctness}). We believe that our findings will provide useful insights for researchers to further improve LLMs for the parallel language translation problem.

UniPar source code and \dataset{} dataset are available at our GitHub
\textcolor{blue}{\href{https://github.com/Scientific-Computing-Lab/UniPar_AI}{repository}}.


\end{abstract}

\section{Introduction}

\begin{figure*}[!htbt]
\centering
\includegraphics[width=0.85\textwidth]{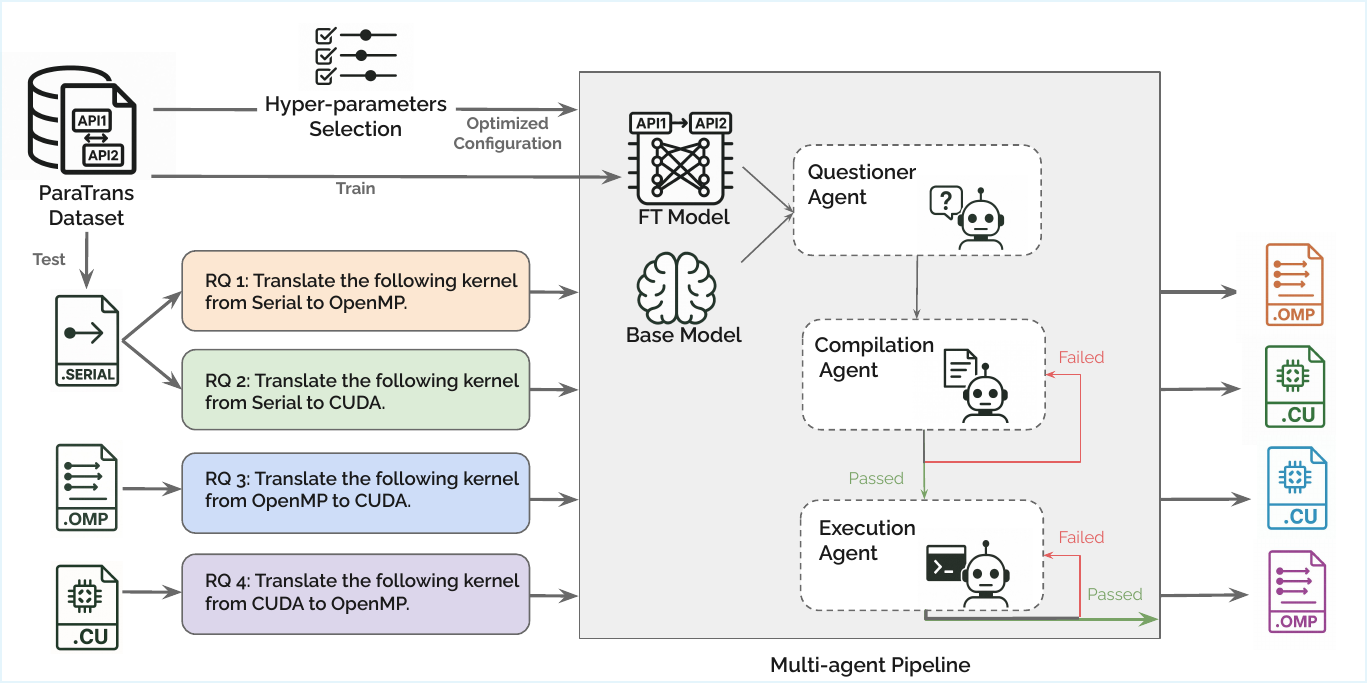}
\caption{
\textbf{UniPar Architecture --- Multi-Agent LLM Pipeline for Parallel Code Translation.}
The system translates kernel functions from the ParaTrans dataset (covering Serial, OpenMP, and CUDA implementations) in one of four directions: Serial$\rightarrow$OpenMP, Serial$\rightarrow$CUDA, OpenMP$\rightarrow$CUDA, or CUDA$\rightarrow$OpenMP. A base or fine-tuned LLM generates candidate translations, optionally using optimized decoding hyperparameters. The multi-agent pipeline then refines these outputs through three components: (1) the \emph{Questioner Agent}, which issues the translation prompt; (2) the \emph{Compilation Agent}, which attempts to compile the generated code and repairs it using compiler diagnostics; and (3) the \emph{Execution Agent}, which validates runtime correctness and triggers additional repair attempts if necessary. This feedback-driven loop continues for up to three rounds. The final output is a syntactically and semantically correct kernel in the desired target API.
}
\label{fig:architecture}
\end{figure*}

Code parallelization is essential across many high-performance computing (HPC) domains, including climate modeling, molecular dynamics, and astrophysics~\cite{hurrell2013community,phillips2005scalable,bryan2014enzo}. To support it, the HPC community and industry have developed various parallel programming languages and frameworks which have seen significantly expanded adoption in recent years~\cite{kadosh2023quantifying} including proprietary options such as CUDA~\cite{nvidia2023cuda} (Nvidia), HIP~\cite{amdHIP} (AMD), as well as open (not proprietary) languages such as OpenMP~\cite{mattson2019openmp}, SYCL~\cite{reyes2016sycl} and OpenCL~\cite{munshi2009opencl}. Proprietary languages, created by hardware vendors to target their devices, create a “lock-in” effect, restricting code to specific hardware. In contrast, open languages offer compatibility across vendors. However, the freedom to easily transfer between these frameworks is highly beneficial in maximizing the best hardware for a task~\cite{pennycook2013investigation}. Whether it be due to features, overall performance, or availability, the ability to easily port code from one parallel computing platform to another grants freedom to researchers and industry alike.

Given the importance of translating between different parallel programming languages, many academic and industry tools have attempted to solve this challenge~\cite{harel2020source,mosseri2020compar}. Recent examples such as the SYCLomatic (CUDA code into SYCL code) and OpenACC-to-OpenMP5 translator from Intel, along with HIPify (CUDA code into HIP code) from AMD, are such tools~\cite{mueller2024syclomatic,servat2022migration,fridman2025openacc,hipify2023}. Academic tools, on the other hand, typically approach the problem as a research problem and offer a wider applicability~\cite{cu2cl2013, kim2012snucl, shi2025transcl}. CU2CL, SnuCL and TransCL are such tools for CUDA to OpenCL translation. Our evaluation of these tools found that static tools quickly become outdated as they are not maintained for updates to compiler tool chains and APIs. Other tools are limited in scope, allowing one to translate exclusively from a single API to a single target API.

More recently, industry-academic efforts such as \cite{mahmud2025contraph,tehrani2024coderosetta, kadosh2023scope, kadosh2023advising, schneider2023mpi,schneider2024mpirigen,harel2023learning, kadosh2024monocoder, chen2024ompgpt} have explored applying machine learning, including small, specialized language models, to address translation tasks between parallel programming APIs. While these approaches offer greater adaptability than static tools in aspects like supporting new APIs or software versions, they still present a relatively high barrier to entry for practical adoption. Motivated by these limitations, we explore an alternative approach that retains the flexibility of learning-based solutions while improving ease of integration and modification as new models become available.

Large language models (LLMs) have recently demonstrated impressive capabilities in code generation and translation between different programming languages~\cite{roziere2020unsupervised,dhruv2025leveraging,tehrani2024coderosetta}, and have been widely studied for their applicability in HPC domains~\cite{chen2024landscape,huang2025paracoder, yadav5114036evaluating, jain2025assessing, yin2025chathpc, mahmud2025autoparllm}. The natural question then is: \emph{How effective are popular LLMs in translating code between different parallel programming languages?} Studies have demonstrated that LLMs exhibit an understanding of parallel programming APIs~\cite{rahman2025performance,bolet2025can,harel2025pragformer}, including work by Nichols et al. \cite{nichols2024can,nichols2024hpc, chaturvedi2025hpc,nichols2025learning} that evaluated LLMs for serial and parallel code generation tasks. Nevertheless, to the best of our knowledge, no existing work evaluates the ability of LLMs to translate code between multiple parallel programming languages with comprehensive techniques to improve their performance.

In this paper, we provide the first systematic evaluation of LLMs (specifically LLaMA-3.3-70B-Instruct and GPT-4o-mini) in their ability to translate between multiple parallel programming languages (namely CUDA and OpenMP). Towards that end, we first develop a cross-paradigm code corpus named \dataset{} and apply it to the baseline evaluation of the LLMs. Additionally, we evaluate four commonly employed improvement strategies built around the LLMs: hyperparameter optimization for decoding, zero- and few-shot prompting, supervised fine-tuning, and iterative feedback through compiler-based repair. Specifically, for every strategy, we evaluate its effects on the model performance and determine if its effects are independent and additive.



\noindent\textbf{Research Questions:} We pose the following RQs:

\begin{itemize}[leftmargin=*,topsep=0pt]
    \item \textbf{RQ1 (Baseline Inference):} Can the base (``vanilla'') instruction-tuned models effectively perform parallel code translation using zero-shot and few-shot prompting (with the best hyperparameters)?
    
    \item \textbf{RQ2 (Supervised Fine-tuning):} Does supervised fine-tuning on the code translation task improve the model’s ability to generate correct and compilable parallel code?
    
    \item \textbf{RQ3 (Agentic Pipeline):} What is the impact of integrating LLMs in an agent-based framework (that provides the compiler feedback) on the translation quality?
\end{itemize}

\noindent\textbf{Contributions:} This paper makes the following contributions:

\begin{enumerate}[leftmargin=*,topsep=0pt]
    \item To the best of our knowledge, UniPar is the first attempt at a systematic evaluation of LLMs on the problem of code translation between parallel programming languages.
    
    \item We construct the first standardized dataset for translation between parallel programming paradigms (Serial, OpenMP, and CUDA), enabling a systematic evaluation.
    
    
    \item We evaluate three different commonly employed strategies to improve the performance of LLMs, namely, few-shot prompting, fine-tuning, and agentic pipelines. The results reveal that although few-shot prompting is ineffective for this problem, both task-specific fine-tuning and agentic framework with compiler feedback deliver considerable improvement to the performance of LLMs, by up to 2X (69\% compilation and 33\% correctness).
\end{enumerate}

\section{Our Approach: UniPar}

We explore three complementary strategies (shown in \autoref{fig:architecture}) to evaluate the ability of an LLM to translate between parallel programming paradigms. First, we assess the foundation models in the zero- and few-shot settings, exploring hyperparameters (e.g., temperature, max token) for optimal performance. Second, we fine-tune an LLM on our curated dataset of aligned kernel implementations, tailoring the model to the translation task. Finally, we develop an agentic pipeline that integrates the LLM with a compiler and an output verifier for an iterative refinement. We used our SOTA cross-paradigm code translation corpus, \dataset{}, to benchmark our methodology (\autoref{sectionparatrans}). 

\input{dataset}
\section{Evaluation}
\label{sec:evaluation}
\subsection{Evaluation Metrics}
To evaluate the capabilities of LLMs in generating parallel code, we adopt a validation methodology inspired by prior work~\cite{kadosh2024ompar, yan2023codescope, ashrafi2025enhancing}, leveraging \hecbench{}'s support for the automated compilation and execution with output verification. Specifically, we evaluate model outputs on the held-out test split of the \dataset{} dataset. This framework enables assessment of both the \emph{compilation rate} -- the percentage of generated programs that compile successfully -- and the \emph{validation rate}, the percentage of generated programs producing outputs that match the expected output specified in the \hecbench{} kernel. For validation, we replaced the \texttt{main} function in the generated code with the \texttt{main} function in the ground-truth kernel, as the \texttt{main} function in the ground-truth kernel contained the output verification check.
In places where the code compiled successfully before the replacement but failed after the replacement, we used 3 attempts with GPT-4o-mini to repair the compilation error. We then verified that no changes were made that affect kernel functionality.   


\subsection{Experimental Setup}

We evaluated the performance of two instruction-tuned foundation models: GPT-4o-mini~\cite{achiam2023gpt} (referred to as GPT in short in the paper) and LLaMA-3.3-70B-Instruct~\cite{grattafiori2024llama} (referred to as LLaMA in short in the paper). These models are selected for their accessibility and strong performance on a wide range of general-purpose code generation tasks, without requiring additional fine-tuning~\cite{chen2021evaluating, hendrycks2020measuring}.


\noindent\textbf{LLM Hyperparameters.} LLMs include several hyperparameters that significantly affect their outputs. The main ones are \textit{temperature}, \textit{top-p}, and \textit{max tokens}. \textit{Temperature} adjusts the randomness (aka ``creativity''): higher values lead to more diverse outputs, while lower values make responses more deterministic. \textit{Top-p} (nucleus sampling) sets a cumulative probability threshold, retaining only the most likely tokens whose total probability exceeds the cutoff. \textit{Max tokens} limits the number of tokens the model may generate in its response (excluding the prompt). After empirical tuning (see~\autoref{subsec:hper_param}), we found that a \textit{temperature} of 0.2 and a \textit{top-p} of 0.9 provided reliable results across both models. We used a maximum token limit of 15,000 whenever possible to accommodate the widest range of model outputs.


\noindent\textbf{LLM Quantization.} To fine-tune the model, we used 16-bit floating point precision, which offers a good balance between computational efficiency and training stability~\cite{micikevicius2017mixed}. For inference, we applied 8-bit quantization to reduce memory usage and improve latency without significantly impacting model performance~\cite{dettmers2023qlora}. 

\noindent\textbf{Hardware.} For LLaMA model inference, we used four NVIDIA L40 GPUs and twelve Intel Xeon Gold 6336Y CPUs. GPT-4o-mini inference was performed via the Azure OpenAI API~\cite{azureopenai}. For the verification evaluation, OpenMP code was executed on a machine equipped with fifty AMD EPYC 7742 CPUs, and CUDA code was run on the same NVIDIA L40 GPUs mentioned above. Fine-tuning was conducted using eight NVIDIA A100 GPUs (80GB).

\section{Results and Analysis} 
\label{chapresults}
In this section, we first answer RQ1, exploring different temperatures 
to present the baseline performance of the chosen LLMs in the zero-shot (\autoref{subsec:hper_param}). Next, we examine the few-shot prompt settings (\autoref{subsec:rq1_few-shot}). We then evaluate the effectiveness of supervised fine-tuning (\autoref{subsec:rq2_finetuning_results}) and the agentic compilation repair pipeline (\autoref{subsec:agent_behaviors}) on the baseline performance of the models.

\subsection{\textbf{RQ1}: Translation Behavior with Zero-shot Prompting}
\label{subsec:hper_param}
We evaluated both LLaMA-3.3-70B-Instruct and GPT-4o-mini across a range of decoding configurations. Specifically, we tested three \textit{temperature} values (0.2, 0.6, and 0.9) and three \textit{max token} limits (5,000, 10,000, and 15,000), while keeping the \textit{top-p} value fixed at 0.8 in a zero-shot prompting setup. Empirically, varying \textit{top-p} did not produce significant improvements in performance, and with a very slight improvement when switching from 0.8 to 0.9. was therefore held constant. 


As shown in ~\autoref{fig:temp_max_token}, lower \textit{temperature} settings yielded both higher and more consistent results across models, with the largest effect observed in the configurations allowing longer outputs. This suggests that reduced randomness in the token selection process improves syntactic validity, especially when generating larger kernel codes, as supported by the findings in previous work~\cite{austin2021program, radford2019language}. This finding is particularly important when accommodating a broader variety of kernels, such as more complex or larger kernels. The rest of our experiments were all done with a temperature of 0.2, top-p of 0.9, and max output tokens set to 15000.

\input{hyper_param_serach_graphs}

\subsection{\textbf{RQ1}: Translation Behavior with Few-shot Prompting}
\label{subsec:rq1_few-shot}

LLMs are trained on a broad range of general tasks. In many scenarios, providing input-output examples during inference (referred to as few-shot prompting) can help guide the model towards the specific task at hand~\cite{bareiss2022code, xu2024does}. To assess its effect in our setting, we compared the compilation rates of generated code with 0-shot, 1-shot, 2-shot, and 3-shot configurations.

Each \emph{n}-shot involved appending \emph{n} additional examples for the same source–target language pair from the \dataset{} train set. We used the prompt addition shown in~\autoref{fig:merged_translation_prompt}. Each example consists of a source kernel and its corresponding translation to the target language, with both the original and the target language specified.

\begin{figure}[ht]
    \centering
    \begin{tcolorbox}[
        width=0.48\textwidth, 
        colback=gray!5, 
        colframe=customgreen, 
        title=Instruction Prompt for Inference,
        fonttitle=\footnotesize, 
        boxsep=1pt,
        top=1.5pt, bottom=1pt
    ]
    \small
    \textbf{System:} You are an HPC expert specializing in translating between parallel programming APIs.
    \hrule
    \vspace{0.3em}
    \textbf{Addition for 1-Shot prompt:}\\
    \textbf{Instruction:}
    Translate the following code from \texttt{\{from\_api\}} to \texttt{\{to\_api\}}\\
    Code: \texttt{\{from\_code\}}\\
    \textbf{Assistant:}
    Here is the translated code: \texttt{\{to\_code\}}
    \hrule
    \vspace{0.3em}
    \textbf{Instruction:}
    Translate the following code from \texttt{\{from\_api\}} to \texttt{\{to\_api\}}\\
    Code: \texttt{\{from\_code\}} \\
    \textbf{Response:} \texttt{\{to\_code\}}
    \end{tcolorbox}
    \caption{
\textbf{Instruction Format for Translation Inference.}
The prompt template used to guide the model during translation between parallel programming schemes, including optional one-shot augmentation with an example input-output pair.
}
    \label{fig:merged_translation_prompt}
\end{figure}

\input{few_shot_graphs}

The results of this experiment are shown in~\autoref{fig:fewshot_comparison}. X axis shows the number of shots provided to the model, while the Y axis shows the compilation rate. The results reveal that both models exhibit minimal sensitivity to the addition of few-shot examples. Compilation success rates remain largely unchanged, except for 3-shot prompting, where performance degrades noticeably. We investigated this behavior further, splitting the task into the topic categories seen in \hecbench{} (e.g., machine learning, math, etc.). We saw that more shots typically did not help the examples in all categories.

While the exact cause of the poor LLM performance with the few-shot prompting would warrant further investigation, specific examples reveal that even on samples similar or identical to the one-shot prompt, LLaMA-3.3-70B-Instruct still introduces subtle but critical syntactic errors. We suspect this results from a combination of context dilution with longer prompts and the nature of few-shot prompting, which is most effective at guiding answers' high-level structure and less at capturing fine-grained details within that structure.
This prompted us to seek a method to more robustly shape the model's domain knowledge, which led us to the experiment with the fine-tuning.


\input{finetunue}

\subsection{\textbf{RQ3}: Agentic Translation with Compiler Feedback}
\label{subsec:agent_behaviors}

To further enhance LLM-based translation, we implemented a multi-agent, feedback-driven architecture, aimed at transforming the translation process from a static process to a dynamic pipeline that mirrors the work of human programmers. The system refines LLM-generated translations through structured iterations grounded in the feedbacks from a compiler and an output verifier. This pipeline, shown in~\autoref{fig:architecture}, is inspired by prior work~\cite{chen2023compcodevet, wang2022compilable, jana2024cotran, chen2024fortran2cpp, dearing2024lassi}, but differs in that we evaluate its utility purely during inference, rather than for the dataset generation.


In our multi-agent pipeline, each agent has access to an LLM model and external tools, including output verifiers and compilers, such as \textit{Clang++} and \textit{NVCC}. Agents rely solely on the current translation status and the available context of the environment for decision-making. The argentic pipeline consists of three agents orchestrated in a feedback cycle:
\begin{itemize}[leftmargin=*,topsep=0pt]
  \item \textbf{Questioner Agent}: Assembles the initial prompts that define the source and target languages (e.g., CUDA → OpenMP). This role can be fulfilled by a vanilla model or its improved version, such as a fine-tuned model.
  \item \textbf{Compilation Agent}: Attempts to compile the generated code. If the compilation fails, it extracts the error message and formulates a compilation repair prompt for its model.
  \item \textbf{Execution Agent}: Executes the code that passes compilation. If output verification fails, it produces an output-repair prompt containing the observed behavior and invokes its model for a correction.
\end{itemize}

Our experiment focused on the use of feedback from tools to improve models' performance. Thus, the pipeline results we report are for a very straightforward pipeline in which the questioner agent first performs an initial translation. Then the compilation agent has three attempts to compile and correct the code according to any errors revealed by the compiler. Finally, if the compilation succeeds, the execution agent has three attempts to run the code on the test cases supplied in the original \hecbench{} code and correct any errors that arise from the feedback from the runtime environment. Note that in this stage, instead of the pass/fail status of the test, we feed the runtime errors for program repair.

In our experiments, we evaluated both LLaMA-3.3-70B-Instruct and GPT-4o-mini as the Questioner agent. Both the Compilation and Execution agents used 3 rounds with GPT-4o-mini as their underlying LLM. All LLMs were configured with a \textit{temperature} of 0.2, \textit{top-p} of 0.9, and a \textit{max token} limit of 15,000. While a number of setup configurations are possible for this experiment, we have reserved that study for future work.


\autoref{fig:agent_pipeline_impact} shows the overall impact of the agentic pipeline on various translation problems, with \autoref{fig:agent_tuple_comparison_compilation} and \autoref{fig:agent_tuple_comparison_validation} showing the performance of the pipeline in terms of compilation and validation rates resp. Specifically, the legend \texttt{Without Agent} refers to the performance without agentic pipeline, while \texttt{With Agent} refers to the performance with the agentic pipeline.
In general, incorporating compiler and output verifier feedback significantly improves overall performance in terms of compilation and validation rate. Repeated rounds of prompting, each incorporating additional diagnostic information, enabled the system to repair a variety of code failures.


\emph{Ablation Study: Effectiveness of different Questioner Models.} We evaluated different Questioner models to understand their impact on the performance (shown in \autoref{fig:agent_comp_val}). We make observation similar to above in this case also --- agentic pipeline has considerable impact on all the Questioner Models. Furthermore, the improvement appears additive, as even when starting from a fine-tuned base model, which already outperforms the vanilla zero-shot baseline, the feedback loop improves performance even further. Although the improvement in the compilation rate is not as pronounced as the improvement in the validation rate, we observe that the fine-tuned model maintains its advantage.

\input{agent_questioner_model}

\emph{Ablation Study: Effectiveness of Compilation Agent in repairing errors.} To analyze the effectiveness of the agentic repair pipeline, we plotted the compilation rate improvements for different repair rounds. \autoref{fig:stacked_pipeline_simple} shows the performance breakdown across different repair rounds and translation problems. We observed that in most of these cases, the first repair attempt provides the largest compilation rate improvement. This stems from the behavior of the model used by the compilation agent that manages to solve only one of two types of issues in each iteration, so codes with multiple needed corrections require more than one iteration. While our current experiments show limited gains from the second iteration, this pattern suggests that longer or more complex code samples may benefit more substantially from additional repair rounds. Furthermore, the gains are not distributed entirely uniformly across translation tasks. Specifically, the pipeline provides slightly larger improvements on the translation problems where the base model initially struggled, such as OMP$\rightarrow$CUDA in LLaMA or Serial$\rightarrow$CUDA on GPT. This equalizing effect suggests that the agent-based system helps surface translation capabilities that are hidden in static prompting setups. The high performance in Serial$\rightarrow$OMP stems from the close similarity between the two, both inherently and due to the generation method described in~\autoref{dataset_preprocessing}.

\input{agent_stack_graphs}


 This result highlights that significant improvements in translation quality can be unlocked via inference time improvements alone, especially when combined with complementary improvements such as fine-tuning.
 
%

\section{Future Work}
Though this preliminary paper performs systematic evaluation of LLMs of their ability to translate between parallel programming languages, we acknowledge that we have not covered a few important problems in this paper. We mention a few of them here. Although we checked the syntax and the functional correctness of the translated code, we did not check its runtime performance or scalability -- factors that are typically important for an HPC application -- nor did we ask LLMs to optimize them. Furthermore, we have reserved an in-depth ablation study along with an evaluation of various design choices for the future. Moreover, current study is restricted to translations to CUDA and OpenMP and its applicability to other important HPC languages such as SYCL, OpenACC, etc., is an interesting question. We will continue to explore some of these problems and encourage the HPC community to join us as well.

\section{conclusion}

In this paper, we introduced UniPar, a unified methodology for generating parallel code as well as translating between a variety of parallel-programming APIs with one system. The framework unifies several enhancements atop a base LLM: (1) exploration of key hyperparameters, including temperature selection and the less effective few-shot inference; (2) fine-tuning on parallel code data; and (3) integration into a dynamic, feedback-driven pipeline. Each component contributes independently to compilation and validation success, with their effects compounding when combined. While UniPar demonstrates significant improvements, further work is needed to enhance compilation and validation rates and to broaden support for longer kernels and a wider range of APIs.

\section*{Acknowledgments}
This research was supported by the Pazy Foundation and the Lynn and William Frankel Center for Computer Science. Computational support was provided by Foundry Cloud Platform and Code Metal.

\bibliographystyle{IEEEtran}
\bibliography{IEEEabrv, references} 

\end{document}

%% file: dataset.tex
\section{\dataset{}: Cross-Paradigm Code Translation Corpus}
\label{sectionparatrans}
To construct a parallel dataset designed to translate between heterogeneous parallel programming languages (e.g., OpenMP and CUDA) and to parallelize serial code, we leveraged the \hecbench{} collection of computing benchmarks~\cite{jin2023benchmark}. \hecbench{} suite comprises 499 benchmarks that cover a wide range of computational domains such as cryptography, machine learning, signal processing, and more. Each benchmark is implemented in multiple parallel programming paradigms, including CUDA, HIP, SYCL/DPC++, and OpenMP-4.5, resulting in 1758 benchmark implementations (as not all benchmarks are available in every parallel programming language). 

\noindent\textbf{Source Scraping.} We applied several filtering criteria to curate the \dataset{} corpus. First, since our focus is exclusively on OpenMP and CUDA, only benchmarks implemented in at least one of these two paradigms were retained. Second, to enable accurate and meaningful translation, it was necessary to isolate the core computational logic of each benchmark, typically encapsulated in a single source file. As our objective is to translate only the core functionality between parallel languages, we analyzed the source code structure of each benchmark.
706 out of 1758 programs consist of a single source file, which clearly contains the primary computational logic. In benchmarks with multiple source files, we observed that auxiliary files such as those named `utils' or `reference' did not contain the main functionality and were therefore excluded. Applying these filters resulted in 406 CUDA-based codes and 276 OpenMP-based codes.

            
\phantomsection
\label{dataset_preprocessing}
\noindent\textbf{Preprocessing.} To standardize the code samples between paradigms, we applied a preprocessing step. First, all comments were removed to reduce noise and focus solely on the code logic. Then, we generated serial versions of the kernels by utilizing the OpenMP-based implementations and removing OpenMP pragmas from them.

\noindent\textbf{Code Verification.} To ensure the quality of the dataset, we applied an automatic validation process to the benchmark kernel codes. This process verifies that each code sample compiles successfully as well as produces outputs consistent with the expected ground-truth results. Kernel implementations that failed this validation step were excluded from the dataset. Note, three of the benchmark kernels did not pass validation and were therefore removed from subsequent experiments.

\noindent\textbf{Token Count Pruning.} Prompting LLMs with entire source codes can result in significant token imbalance across prompts. Prior studies have shown that LLM performance may degrade when processing long or unbalanced inputs, as the context window becomes saturated or loses focus~\cite{liu2023lost}. Moreover, shorter programs often yield more accurate and coherent model output. To address this, we applied a token-length pruning strategy to standardize input lengths and filter out excessively long samples. We set a cut-off of 7,500 tokens, determined empirically by analyzing the token distribution of source files using the LLaMA-3.3-70B-Instruct tokenizer. This threshold effectively balances input coverage and model context fidelity. After applying this constraint, we retained 380 CUDA-based and 258 OpenMP-based programs in \dataset{}.

Altogether, \dataset{} is composed of aligned tuples, with each tuple containing implementations of the same kernel in different parallel programming languages. These tuples enable translation between any pair of languages among Serial, OpenMP, and CUDA. To support a fair evaluation, we split the dataset into a training set and a held-out test set, as shown in \autoref{tab:train_test_split}. The training set contains 898 translation pairs, while the test set comprises 76 pairs, resulting in an approximate ratio of 9 to 1.

\begin{table}[!tbp]
\centering
\small
\caption{The training and test splits of \dataset{}.}
\label{tab:train_test_split}
\begin{minipage}{0.48\linewidth}
\centering
\begin{tabular}{@{}l r@{}}
\toprule
\textbf{Translation} & \textbf{\# Train} \\
\midrule
Serial $\rightarrow$ OpenMP & 235 \\
Serial $\rightarrow$ CUDA & 221 \\
CUDA $\rightarrow$ OpenMP & 221 \\
OpenMP $\rightarrow$ CUDA & 221 \\
\midrule
\textbf{Total} & \textbf{898} \\
\bottomrule
\end{tabular}
\end{minipage}%
\hfill
\begin{minipage}{0.48\linewidth}
\centering
\begin{tabular}{@{}l r@{}}
\toprule
\textbf{Translation} & \textbf{\# Test} \\
\midrule
Serial $\rightarrow$ OpenMP & 20 \\
Serial $\rightarrow$ CUDA & 19 \\
CUDA $\rightarrow$ OpenMP & 18 \\
OpenMP $\rightarrow$ CUDA & 19 \\
\midrule
\textbf{Total} & \textbf{76} \\
\bottomrule
\end{tabular}
\end{minipage}
\end{table}
\vspace{-9pt}

%% file: hyper_param_serach_graphs.tex
\begin{flushleft}
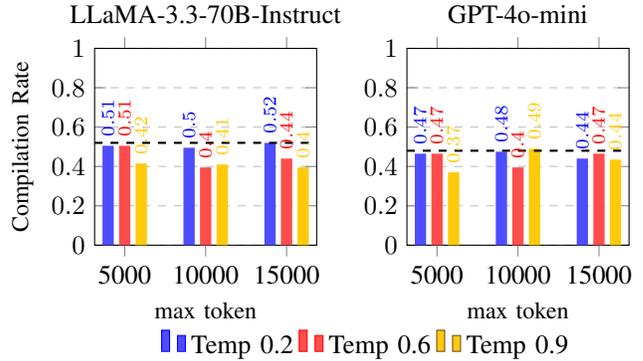
\begin{figure}[!tbp]
\centering
\begin{tikzpicture}
\begin{groupplot}[
    group style={
        group size=2 by 1,
        horizontal sep=1.2cm
    },
    ybar,
    width=0.25\textwidth,%
    height=4.2cm,
    enlarge x limits={abs=0.4cm},
    ylabel style={font=\small},
    xlabel={max token},
    symbolic x coords={5000, 10000, 15000},
    xtick=data,
    ymin=0, ymax=1,
    legend style={anchor=north, legend columns=-1, draw=none, fill=none},
    title={Total Compilation Rate Over Different Temp and Max Tokens (LLAMA3)},
    every node near coord/.append style={font=\small},
    nodes near coords,
every node near coord/.append style={font=\scriptsize, rotate=90, anchor=west},
    ymajorgrids=true,
    grid style=dashed,
    xlabel style={font=\small},
]

\nextgroupplot[
    title={LLaMA-3.3-70B-Instruct},
    bar width=0.15cm, 
    legend style={at={(0.2\textwidth,-0.4)}, anchor=north, legend columns=-1},
     ylabel={Compilation Rate},
]
\addlegendentry{Temp 0.2}
\addlegendentry{Temp 0.6}
\addlegendentry{Temp 0.9}
\addplot+[style={fill=blue!70, draw=none}] plot coordinates {(5000,0.505) (10000,0.495) (15000,0.52)};
\addplot+[style={fill=red!70, draw=none}]  plot coordinates {(5000,0.505) (10000,0.395) (15000,0.44)};
\addplot+[fill=yellow!60!orange, draw=none, every node near coord/.append style={text=yellow!60!orange, draw=none,}] coordinates {(5000,0.415) (10000,0.41) (15000,0.395)};
\draw[dashed, thick] (rel axis cs:0,0.52) -- (rel axis cs:1,0.52);
\nextgroupplot[
    title={GPT-4o-mini},
    bar width=0.15cm, 
]
\addplot+[style={fill=blue!70, draw=none}] plot coordinates {(5000,0.465) (10000,0.475) (15000,0.44)};
\addplot+[style={fill=red!70, draw=none}]  plot coordinates {(5000,0.465) (10000,0.395) (15000,0.465)};
\addplot+[fill=yellow!60!orange, draw=none, every node near coord/.append style={text=yellow!60!orange, draw=none,}] coordinates {(5000,0.37) (10000,0.49) (15000,0.435)};
\draw[dashed, thick] (rel axis cs:0,0.48) -- (rel axis cs:1,0.48);
\end{groupplot}
\end{tikzpicture}
\caption{
\textbf{Effect of Temperature and Max Token Settings on Compilation Rate.}
Compilation success rates for the models across combinations of temperature and maximum token limits. Lower temperatures, particularly 0.2, yield the highest and most consistent compilation rates, with the effect especially pronounced at higher token lengths for LLaMA.
}
\label{fig:temp_max_token}
\end{figure}
\end{flushleft}

%% file: few_shot_graphs.tex
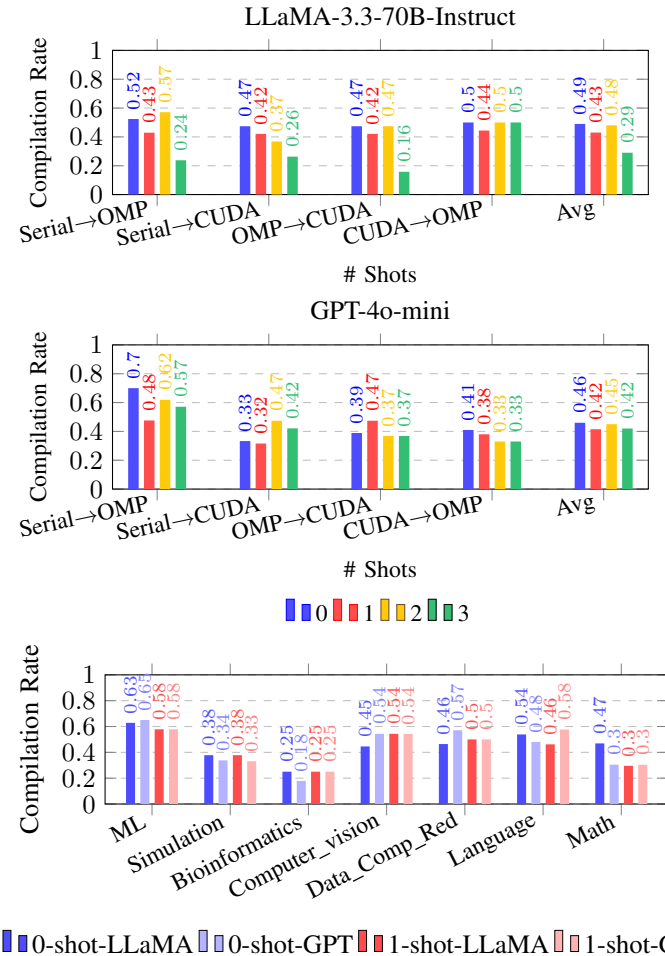
\begin{figure}[!tbp]
\centering

\begin{tikzpicture}
\begin{groupplot}[
    group style={
        group size=1 by 2, 
        horizontal sep=0.8cm,
        vertical sep=2cm
    },
    ybar,
    width=0.48\textwidth, 
    height=3.5cm,
    ymin=0, ymax=1,
    ymajorgrids=true,
    grid style=dashed,
    ylabel style={font=\small},
    every node near coord/.append style={font=\small},
    nodes near coords,
    every node near coord/.append style={font=\scriptsize, rotate=90, anchor=west},
    xtick=data,
    xlabel={\# Shots},
    xlabel style={font=\small}
]
\nextgroupplot[
    title={LLaMA-3.3-70B-Instruct},
    bar width=0.14cm,
    symbolic x coords={Serial→OMP, Serial→CUDA, OMP→CUDA, CUDA→OMP, Avg},
    ylabel={Compilation Rate},
    xticklabel style={rotate=12, anchor=east, font=\small},
]
\addplot+[style={fill=blue!70,draw=none}] plot coordinates {(Serial→OMP,0.524) (Serial→CUDA,0.474) (OMP→CUDA,0.474) (CUDA→OMP,0.5) (Avg, 0.49)};
\addplot+[style={fill=red!70,draw=none}]  plot coordinates {(Serial→OMP,0.429) (Serial→CUDA,0.421) (OMP→CUDA,0.421) (CUDA→OMP,0.444) (Avg, 0.43)};
\addplot+[fill=yellow!60!orange, draw=none, every node near coord/.append style={text=yellow!60!orange, draw=none,}] coordinates {(Serial→OMP,0.571) (Serial→CUDA,0.368) (OMP→CUDA,0.474) (CUDA→OMP,0.5) (Avg, 0.48)};
\addplot+[fill=green!68!blue!80,draw=none, every node near coord/.append style={text=green!68!blue!80, draw=none,}] coordinates {(Serial→OMP,0.238) (Serial→CUDA,0.263) (OMP→CUDA,0.158) (CUDA→OMP,0.5) (Avg, 0.29)};

\nextgroupplot[
    title={GPT-4o-mini},
    bar width=0.14cm,
    symbolic x coords={Serial→OMP, Serial→CUDA, OMP→CUDA, CUDA→OMP, Avg}, 
    legend style={at={(0.5,-0.72)}, font=\small, anchor=north, legend columns=-1, draw=none, fill=none},
    xticklabel style={rotate=12, anchor=east, font=\small},
    ylabel={Compilation Rate},
]
\legend{0,1,2,3} 
\addplot+[style={fill=blue!70,draw=none}] plot coordinates {(Serial→OMP,0.7) (Serial→CUDA,0.333) (OMP→CUDA,0.389) (CUDA→OMP,0.41) (Avg, 0.46)};
\addplot+[style={fill=red!70,draw=none}]  plot coordinates {(Serial→OMP,0.476) (Serial→CUDA,0.316) (OMP→CUDA,0.474) (CUDA→OMP,0.38) (Avg, 0.415)};

\addplot+[fill=yellow!60!orange, draw=none, every node near coord/.append style={text=yellow!60!orange, draw=none,}] coordinates {(Serial→OMP,0.619) (Serial→CUDA,0.474) (OMP→CUDA,0.368) (CUDA→OMP,0.33) (Avg, 0.45)};

\addplot+[fill=green!68!blue!80,draw=none, every node near coord/.append style={text=green!68!blue!80, draw=none,}] coordinates {(Serial→OMP,0.571) (Serial→CUDA,0.421) (OMP→CUDA,0.368) (CUDA→OMP,0.33) (Avg, 0.42)};

\end{groupplot}
\end{tikzpicture}

\begin{tikzpicture}
\begin{axis}[
    ybar,
    width=0.5\textwidth, 
    height=3.3cm,
    bar width=0.12cm,
    ylabel={Compilation Rate},
    symbolic x coords={ML, Simulation, Bioinformatics, Computer\_vision, Data\_Comp\_Red, Language, Math},
    xtick=data,
    nodes near coords,
    every node near coord/.append style={font=\scriptsize, rotate=90, anchor=west, draw=none},
    xticklabel style={rotate=27, anchor=east, font=\small},
    ymin=0, ymax=1,
    ymajorgrids=true,
    grid style=dashed,
    legend style={at={(0.45,-0.93)}, anchor=north, legend columns=-1, draw=none, fill=none},
]

\pgfplotsset{
    custom legend 2/.style={
        legend image post style={
            draw=black, 
            line width=0.1pt,
        }
    }
}

\addplot+[fill=blue!70, draw=none,  custom legend 2, every node near coord/.append style={text=blue!70,
        draw=none,}] coordinates { (ML,0.62825) (Simulation,0.378) (Bioinformatics,0.25025) (Computer\_vision,0.44575) (Data\_Comp\_Red,0.46425) (Language,0.5385) (Math,0.46925) };
\addplot+[fill=blue!30, draw=none, custom legend 2, every node near coord/.append style={text=blue!50,
        draw=none,}] coordinates { (ML,0.65025) (Simulation,0.33775) (Bioinformatics,0.17875) (Computer\_vision,0.5435) (Data\_Comp\_Red,0.57125) (Language,0.48075) (Math,0.3045) };

\addplot+[fill=red!70, draw=none,  custom legend 2, every node near coord/.append style={text=red!70,
        draw=none,}] coordinates { (ML,0.5785) (Simulation,0.378) (Bioinformatics,0.25025) (Computer\_vision,0.5435) (Data\_Comp\_Red,0.5) (Language,0.462) (Math,0.2955) };
\addplot+[fill=red!30, draw=none,  custom legend 2, every node near coord/.append style={text=red!50,
        draw=none,}] coordinates { (ML,0.5785) (Simulation,0.331) (Bioinformatics,0.25025) (Computer\_vision,0.54325) (Data\_Comp\_Red,0.5) (Language,0.5765) (Math,0.304) };

\legend{0-shot-LLaMA, 0-shot-GPT, 1-shot-LLaMA, 1-shot-GPT}
\end{axis}
\end{tikzpicture}
\caption{
\textbf{Effect of Few-Shot Prompting on Compilation Rate.}
Compilation rates for the models across 0- to 3-shot configurations and for different \hecbench{} categories. Increasing the number of prompt examples did not consistently improve the compilation rate and can lead to degradation, especially beyond 1-shot. Effects vary by translation direction and application domain.
}

\label{fig:fewshot_comparison}
\end{figure}

%% file: finetunue.tex
\subsection{\textbf{RQ2}: Supervised Fine-Tuning}
\label{subsec:rq2_finetuning_results}
Base instruction-tuned models such as GPT and LLaMA are trained for a general use. To adapt these models specifically for the task of parallel code translation, we fine-tuned them on the \dataset{} training set to narrow their scope and enhance their performance~\cite{kadosh2023monocoder, kadosh2024ompar, schneider2024mpirigen, chaturvedi2024hpc}.

We used the instruction-tuned version of LLaMA-3.3-70B-Instruct model, leveraging the alignment already present to accelerate convergence and improve performance, as instruction-tuned models require less supervision to adapt to new tasks compared to fine-tuning from a base model~\cite{ shengyu2023instruction,wu2025shadow}. We trained the model on aligned kernel tuples from \dataset{}. The prompt format used in training is shown in~\autoref{fig:merged_translation_prompt}.


 To accommodate source codes spanning thousands of tokens, we maximized the model’s context length within the hardware memory constraints. Using memory optimization techniques provided by the TorchTune library~\cite{torchtune}, we configured the model to operate with the context length of 16,384 tokens.

\input{finetune_result_graph}

\input{agent_graph_comparison}

\autoref{fig:finetune_tuple_comparison} shows the performance of the base and the fine-tuned LLaMA model across all translation problems on the held-out test set of the \dataset{} corpus. The results reveal that the fine-tuned model shows a performance that is on par with or exceeds that of the base models in terms of the compilation rate.
In particular, the fine-tuned model outperforms LLaMA-3.3-70B-Instruct in generating semantically valid code, achieving a validation rate of 29.4\% for translating to CUDA and 38.2\% for translating to OpenMP.

%% file: finetune_result_graph.tex
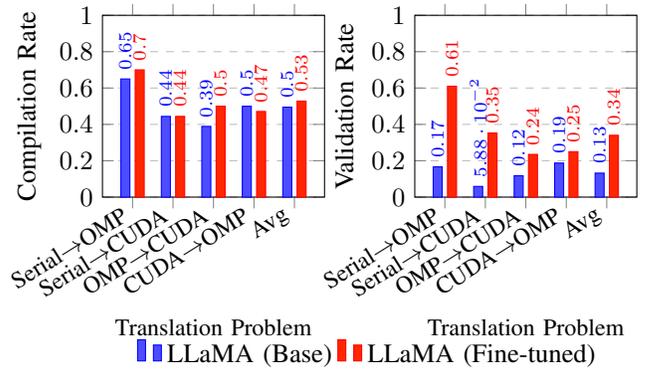
\begin{figure}[!tbp]
\centering

\centering
\begin{tikzpicture}
\begin{groupplot}[
    group style={
        group size=2 by 1,
        horizontal sep=1.2cm,
    },
    ybar,
    width=0.25\textwidth,
    height=4.0cm,
    enlarge x limits={abs=0.4cm},
    xtick=data,
    xticklabel style={rotate=32, anchor=east, font=\small},
    symbolic x coords={Serial$\rightarrow$OMP, Serial$\rightarrow$CUDA, OMP$\rightarrow$CUDA, CUDA$\rightarrow$OMP, Avg},
    nodes near coords,
every node near coord/.append style={font=\scriptsize, rotate=90, anchor=west},
    xlabel={Translation Problem},
    ymin=0, ymax=1,
    ymajorgrids=true,
    grid style=dashed,
    legend style={at={(1.2,-0.75)}, anchor=north, legend columns=-1, draw=none, fill=none},
    xlabel style={font=\small},
]
\nextgroupplot[bar width=0.12cm,  ylabel={Compilation Rate},]

\addplot+[fill=blue!70] coordinates {(Serial$\rightarrow$OMP,0.65) (Serial$\rightarrow$CUDA,0.444) (OMP$\rightarrow$CUDA,0.389) (CUDA$\rightarrow$OMP,0.5) (Avg,0.495)};
\addplot+[fill=red!70!orange] coordinates {(Serial$\rightarrow$OMP,0.7) (Serial$\rightarrow$CUDA,0.444) (OMP$\rightarrow$CUDA,0.5) (CUDA$\rightarrow$OMP,0.471) (Avg, 0.528)};

\legend{LLaMA (Base), LLaMA (Fine-tuned)}

\nextgroupplot[bar width=0.12cm,  ylabel={Validation Rate},]

\addplot+[fill=blue!70] coordinates {(Serial$\rightarrow$OMP,0.16666) (Serial$\rightarrow$CUDA,0.0588) (OMP$\rightarrow$CUDA,0.1176) (CUDA$\rightarrow$OMP,0.1875) (Avg, 0.132) };
\addplot+[fill=red!70!orange] coordinates {(Serial$\rightarrow$OMP,0.61) (Serial$\rightarrow$CUDA,0.3529) (OMP$\rightarrow$CUDA,0.235) (CUDA$\rightarrow$OMP,0.25) (Avg, 0.34)};

\end{groupplot}
\end{tikzpicture}

\caption{
\textbf{Effect of Fine-Tuning on Compilation and Validation Rates.}
Compilation and validation rates for the LLaMA model across translation directions before and after fine-tuning. While compilation rate improvements are modest, fine-tuning substantially increases the proportion of functionally-correct outputs, with average validation rates more than doubling.
}
\label{fig:finetune_tuple_comparison}\end{figure}

%% file: agent_graph_comparison.tex
\begin{figure*}[!ht]
\centering
\begin{subfigure}{0.48\textwidth}
\centering
\begin{tikzpicture}
\begin{groupplot}[
    group style={
        group size=2 by 1,
        horizontal sep=1.2cm,
    },
    ybar,
    height=3.9cm,
    enlarge x limits={abs=0.4cm},
    xtick=data,
    xticklabel style={rotate=32, anchor=east, font=\small},
    symbolic x coords={Serial$\rightarrow$OMP, Serial$\rightarrow$CUDA, OMP$\rightarrow$CUDA, CUDA$\rightarrow$OMP, Avg},
    nodes near coords,
every node near coord/.append style={font=\scriptsize, rotate=90, anchor=west},
    xlabel={Translation Problem},
    ymin=0, ymax=1,
    ymajorgrids=true,
    grid style=dashed,
    legend style={at={(1.2,-1.0)}, anchor=north, legend columns=-1, draw=none, fill=none},
    xlabel style={font=\small},
]
\nextgroupplot[title={LLaMA-3.3-70B-Instruct}, bar width=0.12cm,  ylabel={Compilation Rate},]

\addplot+[fill=blue!70] coordinates {(Serial$\rightarrow$OMP,0.65) (Serial$\rightarrow$CUDA,0.444) (OMP$\rightarrow$CUDA,0.389) (CUDA$\rightarrow$OMP,0.529) (Avg, 0.50)};

\addplot+[fill=red!70!orange] coordinates {(Serial$\rightarrow$OMP,0.8) (Serial$\rightarrow$CUDA,0.667) (OMP$\rightarrow$CUDA,0.722) (CUDA$\rightarrow$OMP,0.765) (Avg, 0.735)};

\legend{Without agent, With agent}

\nextgroupplot[title={GPT-4o-mini}, bar width=0.12cm,]

\addplot+[fill=blue!70] coordinates {(Serial$\rightarrow$OMP,0.7) (Serial$\rightarrow$CUDA,0.333) (OMP$\rightarrow$CUDA,0.389) (CUDA$\rightarrow$OMP,0.412) (Avg, 0.46) };

\addplot+[fill=red!70!orange] coordinates {(Serial$\rightarrow$OMP,0.95) (Serial$\rightarrow$CUDA,0.611) (OMP$\rightarrow$CUDA,0.556) (CUDA$\rightarrow$OMP,0.647) (Avg, 0.69)};

\end{groupplot}
\end{tikzpicture}
\caption{
\textbf{Impact of Agentic Pipeline on Compilation Rate Across Translation Problems.}
Compilation rates for Questioner LLMs across all translation problems, with and without the agentic pipeline (Compilation and Execution agents). Agent-based repair has a slight equalizing effect on results across translation directions as the variation in compilation rate shrinks.
}
\label{fig:agent_tuple_comparison_compilation}
\end{subfigure}
\hfill
\begin{subfigure}{0.48\textwidth}
\centering

\centering
\begin{tikzpicture}
\begin{groupplot}[
    group style={
        group size=2 by 1,
        horizontal sep=1.2cm,
    },
    ybar,
    height=3.9cm,
    enlarge x limits={abs=0.4cm},
    xtick=data,
    xticklabel style={rotate=32, anchor=east, font=\small},
    symbolic x coords={Serial$\rightarrow$OMP, Serial$\rightarrow$CUDA, OMP$\rightarrow$CUDA, CUDA$\rightarrow$OMP, Avg},
    nodes near coords,
every node near coord/.append style={font=\scriptsize, rotate=90, anchor=west},
    xlabel={Translation Problem},
    ymin=0, ymax=1,
    ymajorgrids=true,
    grid style=dashed,
    legend style={at={(1.2,-1.0)}, anchor=north, legend columns=-1, draw=none, fill=none},
    xlabel style={font=\small},
]
\nextgroupplot[title={LLaMA-3.3-70B-Instruct}, bar width=0.12cm,  ylabel={Validation Rate},]

\addplot+[fill=blue!70] coordinates {(Serial$\rightarrow$OMP,0.1666) (Serial$\rightarrow$CUDA,0.0588) (OMP$\rightarrow$CUDA,0.1176) (CUDA$\rightarrow$OMP,0.1875) (Avg, 0.1324)};

\addplot+[fill=red!70!orange] coordinates {(Serial$\rightarrow$OMP,0.33333) (Serial$\rightarrow$CUDA,0.3529) (OMP$\rightarrow$CUDA,0.3823) (CUDA$\rightarrow$OMP,0.25) (Avg, 0.327)};

\legend{Without agent, With agent}

\nextgroupplot[title={GPT-4o-mini}, bar width=0.12cm,]

\addplot+[fill=blue!70] coordinates {(Serial$\rightarrow$OMP,0.277777) (Serial$\rightarrow$CUDA,0.088) (OMP$\rightarrow$CUDA,0.0588) (CUDA$\rightarrow$OMP,0.1875) (Avg, 0.152) };

\addplot+[fill=red!70!orange] coordinates {(Serial$\rightarrow$OMP,0.33333) (Serial$\rightarrow$CUDA,0.4) (OMP$\rightarrow$CUDA,0.29411) (CUDA$\rightarrow$OMP,0.3125) (Avg, 0.334)};

\end{groupplot}
\end{tikzpicture}
\caption{
\textbf{Impact of Agentic Pipeline on Validation Rate Across Translation Problems.}
Validation rates for Questioner LLMs across all translation problems, with and without the agentic pipeline (Compilation and Execution agents). Agent-based repair substantially improves functional correctness in every direction, including challenging cases such as OMP$\rightarrow$CUDA and Serial$\rightarrow$CUDA.
}
\label{fig:agent_tuple_comparison_validation}
\end{subfigure}
\vspace{0.3cm}
\caption{\textbf{Impact of Agentic Pipeline}}
\label{fig:agent_pipeline_impact}
\end{figure*}
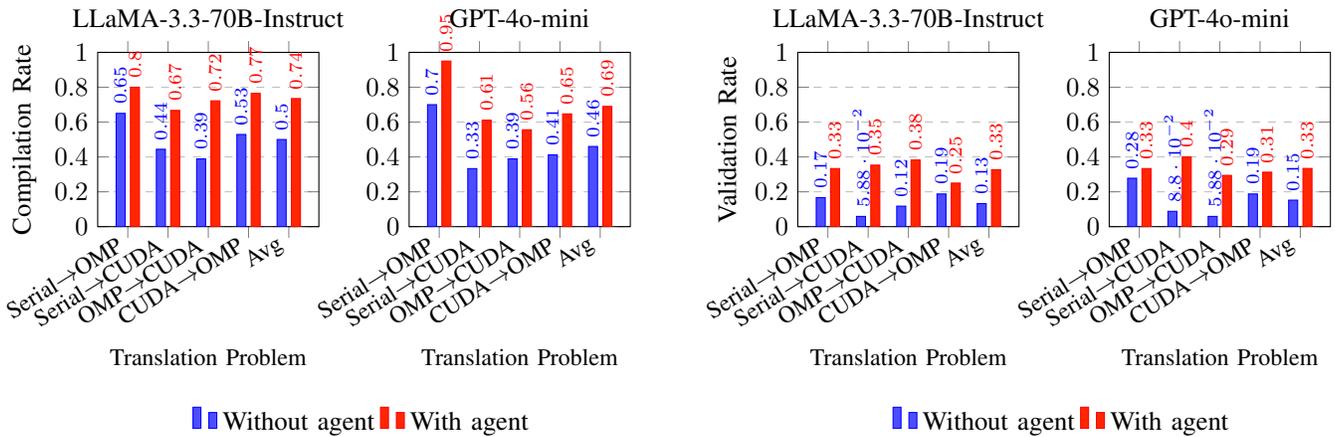


%% file: agent_questioner_model.tex
\begin{figure}[!tbp]
\centering
\begin{tikzpicture}
\begin{groupplot}[
    group style={
        group size=2 by 1,
        horizontal sep=1.2cm
    },
    ybar,
    width=0.25\textwidth,
    height=4.2cm,
    enlarge x limits={abs=0.4cm},
    symbolic x coords={LLaMA (Base), GPT-4o-mini (Base), LLaMA (FT)},
    nodes near coords,
every node near coord/.append style={font=\scriptsize, rotate=90, anchor=west},
    xtick=data,
    xticklabel style={rotate=22, anchor=east, font=\small},
    ymin=0, ymax=1,
    ymajorgrids=true,
    grid style=dashed,
    legend style={at={(1.1,-0.7)}, anchor=north, legend columns=-1, draw=none, fill=none},
    xlabel style={font=\small},
]

\nextgroupplot[
    ylabel={Compilation Rate},
    xlabel={Questioner Model},
    bar width=0.12cm,
]
\addplot+[fill=blue!70] coordinates {(LLaMA (FT),0.53) (LLaMA (Base),0.495) (GPT-4o-mini (Base),0.46)};

\addplot+[fill=red!70] coordinates {(LLaMA (FT),0.695) (LLaMA (Base),0.735) (GPT-4o-mini (Base),0.69)};

\addlegendentry{Without agent}
\addlegendentry{With agent}

\nextgroupplot[
    ylabel={Validation Rate},
    xlabel={Questioner Model},
    bar width=0.12cm,
]
\addplot+[fill=blue!70] coordinates {(LLaMA (FT),0.338) (LLaMA (Base),0.176) (GPT-4o-mini (Base),0.1545)};
\addplot+[fill=red!70] coordinates {(LLaMA (FT),0.397) (LLaMA (Base),0.331) (GPT-4o-mini (Base),0.3385)};

\end{groupplot}
\end{tikzpicture}

\caption{
\textbf{Overall Impact of the Agentic Pipeline Across Different Questioner Models.}
Compilation and validation rates for various Questioner models with and without the agentic pipeline (Compilation and Execution agents). Incorporating Compilation and Execution agents leads to a substantial improvement in both metrics, especially for base models. While the fine-tuned model shows smaller gains, it consistently achieves the highest overall validation rate.
}

\label{fig:agent_comp_val}
\end{figure}
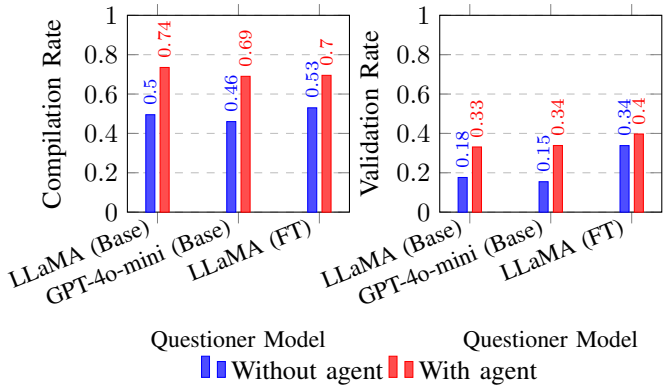

%% file: agent_stack_graphs.tex
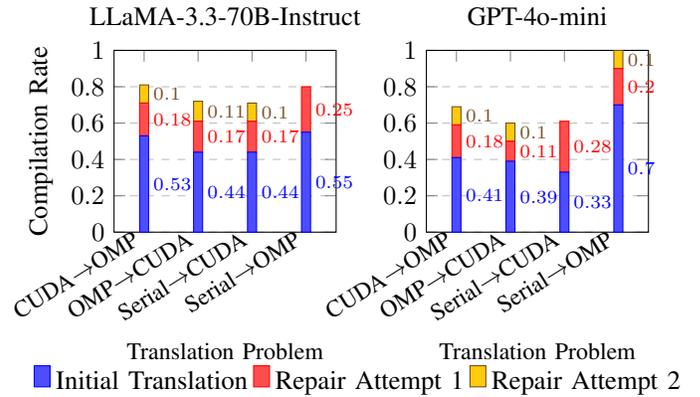
\begin{figure}[!tbp]
\centering
\begin{tikzpicture}
\begin{groupplot}[
    group style={
        group size=2 by 1,
        horizontal sep=1.2cm,
    },
    ybar stacked,
    width=0.25\textwidth,
    height=4.0cm,
    ymin=0, ymax=1,
    enlarge x limits={abs=0.4cm},
    xlabel={Translation Problem},
    symbolic x coords={CUDA$\rightarrow$OMP, OMP$\rightarrow$CUDA, Serial$\rightarrow$CUDA, Serial$\rightarrow$OMP},
    nodes near coords,
every node near coord/.append style={font=\scriptsize, rotate=0, anchor=west},
    xtick=data,
    xticklabel style={rotate=32, anchor=east, font=\small},
    ymajorgrids=true,
    grid style=dashed,
    legend style={at={(-0.3,-0.7)}, anchor=north, legend columns=-1, draw=none, fill=none},
    xlabel style={font=\small},
    ]

\nextgroupplot[title={LLaMA-3.3-70B-Instruct}, bar width=0.12cm,  ylabel={Compilation Rate},]


\addplot+[fill=blue!70]  coordinates {(CUDA$\rightarrow$OMP,0.53) (OMP$\rightarrow$CUDA,0.44) (Serial$\rightarrow$CUDA,0.44) (Serial$\rightarrow$OMP,0.55)};

\addplot+[fill=red!70]   coordinates {(CUDA$\rightarrow$OMP,0.18) (OMP$\rightarrow$CUDA,0.17) (Serial$\rightarrow$CUDA,0.17) (Serial$\rightarrow$OMP,0.25)};

\addplot+[fill=yellow!60!orange] coordinates {(CUDA$\rightarrow$OMP,0.1) (OMP$\rightarrow$CUDA,0.11) (Serial$\rightarrow$CUDA,0.1) (Serial$\rightarrow$OMP,0)};

\nextgroupplot[title={GPT-4o-mini}, bar width=0.12cm]

\legend{Initial Translation, Repair Attempt 1, Repair Attempt 2}
\addplot+[fill=blue!70]  coordinates {(CUDA$\rightarrow$OMP,0.41) (OMP$\rightarrow$CUDA,0.39) (Serial$\rightarrow$CUDA,0.33) (Serial$\rightarrow$OMP,0.7)};
\addplot+[fill=red!70]   coordinates {(CUDA$\rightarrow$OMP,0.18) (OMP$\rightarrow$CUDA,0.11) (Serial$\rightarrow$CUDA,0.28) (Serial$\rightarrow$OMP,0.2)};
\addplot+[fill=yellow!60!orange] coordinates {(CUDA$\rightarrow$OMP,0.1) (OMP$\rightarrow$CUDA,0.1) (Serial$\rightarrow$CUDA,0.0) (Serial$\rightarrow$OMP,0.1)};

\end{groupplot}
\end{tikzpicture}
\caption{
\textbf{Effect of Compilation Agent's Repair Attempts on Compilation Rate.}
Compilation rates for the models across translation directions, broken down by initial generation and up to two rounds of agent-guided repair. The first repair attempt yields the largest improvement, particularly in directions with initially low success rates. Additional gains from the second repair attempt help equalize performance across translation tuples.
}
\label{fig:stacked_pipeline_simple}
\end{figure}

%% file: main.bbl
\begin{thebibliography}{10}
\providecommand{\url}[1]{#1}
\csname url@samestyle\endcsname
\providecommand{\newblock}{\relax}
\providecommand{\bibinfo}[2]{#2}
\providecommand{\BIBentrySTDinterwordspacing}{\spaceskip=0pt\relax}
\providecommand{\BIBentryALTinterwordstretchfactor}{4}
\providecommand{\BIBentryALTinterwordspacing}{\spaceskip=\fontdimen2\font plus
\BIBentryALTinterwordstretchfactor\fontdimen3\font minus \fontdimen4\font\relax}
\providecommand{\BIBforeignlanguage}[2]{{%
\expandafter\ifx\csname l@#1\endcsname\relax
\typeout{** WARNING: IEEEtran.bst: No hyphenation pattern has been}%
\typeout{** loaded for the language `#1'. Using the pattern for}%
\typeout{** the default language instead.}%
\else
\language=\csname l@#1\endcsname
\fi
#2}}
\providecommand{\BIBdecl}{\relax}
\BIBdecl

\bibitem{hurrell2013community}
J.~W. Hurrell, M.~M. Holland, P.~R. Gent, S.~Ghan, J.~E. Kay, P.~J. Kushner, J.-F. Lamarque, W.~G. Large, D.~Lawrence, K.~Lindsay \emph{et~al.}, ``The community earth system model: a framework for collaborative research,'' \emph{Bulletin of the American Meteorological Society}, vol.~94, no.~9, pp. 1339--1360, 2013.

\bibitem{phillips2005scalable}
J.~C. Phillips, R.~Braun, W.~Wang, J.~Gumbart, E.~Tajkhorshid, E.~Villa, C.~Chipot, R.~D. Skeel, L.~Kale, and K.~Schulten, ``Scalable molecular dynamics with namd,'' \emph{Journal of computational chemistry}, vol.~26, no.~16, pp. 1781--1802, 2005.

\bibitem{bryan2014enzo}
G.~L. Bryan, M.~L. Norman, B.~W. O'Shea, T.~Abel, J.~H. Wise, M.~J. Turk, D.~R. Reynolds, D.~C. Collins, P.~Wang, S.~W. Skillman \emph{et~al.}, ``Enzo: An adaptive mesh refinement code for astrophysics,'' \emph{The Astrophysical Journal Supplement Series}, vol. 211, no.~2, p.~19, 2014.

\bibitem{kadosh2023quantifying}
T.~Kadosh, N.~Hasabnis, T.~Mattson, Y.~Pinter, and G.~Oren, ``Quantifying openmp: Statistical insights into usage and adoption,'' 2023.

\bibitem{nvidia2023cuda}
``{CUDA C++ Programming Guide},'' \url{https://docs.nvidia.com/cuda/cuda-c-programming-guide/index.html}, 2023, [Online].

\bibitem{amdHIP}
{ROCm Developer Tools Team}, ``{HIP}: Heterogeneous-compute interface for portability,'' \url{https://github.com/ROCm-Developer-Tools/HIP}, 2025, accessed: 2025-06-09.

\bibitem{mattson2019openmp}
T.~Mattson, Y.~H. He, and A.~Koniges, \emph{{The OpenMP Common Core: Making OpenMP Simple Again (Scientific and Engineering Computation)}}.\hskip 1em plus 0.5em minus 0.4em\relax {The MIT Press}, 2019.

\bibitem{reyes2016sycl}
R.~Reyes and V.~Lom{\"u}ller, ``{SYCL: Single-source C++ accelerator programming},'' in \emph{Parallel Computing: On the Road to Exascale}.\hskip 1em plus 0.5em minus 0.4em\relax IOS Press, 2016.

\bibitem{munshi2009opencl}
A.~Munshi, ``The opencl specification,'' in \emph{2009 IEEE Hot Chips 21 Symposium (HCS)}.\hskip 1em plus 0.5em minus 0.4em\relax IEEE, 2009, pp. 1--314.

\bibitem{pennycook2013investigation}
S.~J. Pennycook, S.~D. Hammond, S.~A. Wright, J.~Herdman, I.~Miller, and S.~A. Jarvis, ``An investigation of the performance portability of opencl,'' \emph{Journal of Parallel and Distributed Computing}, vol.~73, no.~11, pp. 1439--1450, 2013.

\bibitem{harel2020source}
R.~Harel, I.~Mosseri, H.~Levin, L.-o. Alon, M.~Rusanovsky, and G.~Oren, ``Source-to-source parallelization compilers for scientific shared-memory multi-core and accelerated multiprocessing: analysis, pitfalls, enhancement and potential,'' \emph{International Journal of Parallel Programming}, vol.~48, pp. 1--31, 2020.

\bibitem{mosseri2020compar}
I.~Mosseri, L.-o. Alon, R.~Harel, and G.~Oren, ``Compar: optimized multi-compiler for automatic openmp s2s parallelization,'' in \emph{OpenMP: Portable Multi-Level Parallelism on Modern Systems: 16th International Workshop on OpenMP, IWOMP 2020, Austin, TX, USA, September 22--24, 2020, Proceedings 16}.\hskip 1em plus 0.5em minus 0.4em\relax Springer, 2020, pp. 247--262.

\bibitem{mueller2024syclomatic}
R.~Mueller-Albrecht, ``Syclomatic: Sycl adoption for everyone-moving from cuda to sycl gets progressively easier: Advanced migration considerations,'' in \emph{Proceedings of the 12th International Workshop on OpenCL and SYCL}, 2024, pp. 1--2.

\bibitem{servat2022migration}
H.~Servat, G.~Rossi, A.~Duran, and R.~Narayanaswamy, ``On the migration of openacc-based applications into openmp 5+,'' in \emph{International Workshop on OpenMP}.\hskip 1em plus 0.5em minus 0.4em\relax Springer, 2022, pp. 127--141.

\bibitem{fridman2025openacc}
Y.~Fridman, Y.~Goren, and G.~Oren, ``From openacc to openmp5 gpu offloading: Performance evaluation on nas parallel benchmarks,'' in \emph{Proceedings of the 2025 4th International Workshop on Extreme Heterogeneity Solutions}, 2025, pp. 10--18.

\bibitem{hipify2023}
{ROCm Developers}, ``{HIPIFY}: Cuda to hip source translation tool,'' \url{https://github.com/ROCm/HIPIFY}, 2023, accessed: 2025-06-09.

\bibitem{cu2cl2013}
A.~Head, ``{cu2cl}: Cuda to opencl source-to-source translator,'' \url{https://github.com/andrewhead/cu2cl}, 2013, accessed: 2025-06-09.

\bibitem{kim2012snucl}
J.~Kim, S.~Seo, J.~Lee, J.~Nah, G.~Jo, and J.~Lee, ``Snucl: an opencl framework for heterogeneous cpu/gpu clusters,'' in \emph{Proceedings of the 26th ACM international conference on Supercomputing}, 2012, pp. 341--352.

\bibitem{shi2025transcl}
C.~Shi, Y.~Sun, R.~Chen, J.~Wang, Q.~Guo, C.~Gong, Y.~Sui, Y.~Jin, and Y.~Zhang, ``Transcl: An automatic cuda-to-opencl programs transformation framework,'' \emph{ACM Transactions on Architecture and Code Optimization}, 2025.

\bibitem{mahmud2025contraph}
Q.~I. Mahmud, A.~TehraniJamsaz, N.~K. Ahmed, T.~L. Willke, and A.~Jannesari, ``Contraph: Contrastive learning for parallelization and performance optimization,'' \emph{Proceedings of the 2025 ACM International Conference on Supercomputing (ICS ’25)}, 2025.

\bibitem{tehrani2024coderosetta}
A.~Tehrani, A.~Bhattacharjee, L.~Chen, N.~K. Ahmed, A.~Yazdanbakhsh, and A.~Jannesari, ``Coderosetta: Pushing the boundaries of unsupervised code translation for parallel programming,'' \emph{Advances in Neural Information Processing Systems}, vol.~37, pp. 100\,965--100\,999, 2024.

\bibitem{kadosh2023scope}
T.~Kadosh, N.~Hasabnis, V.~A. Vo, N.~Schneider, N.~Krien, A.~Wasay, N.~Ahmed, T.~Willke, G.~Tamir, Y.~Pinter \emph{et~al.}, ``{Scope is all you need: Transforming LLMs for HPC Code},'' \emph{arXiv preprint arXiv:2308.09440}, 2023.

\bibitem{kadosh2023advising}
T.~Kadosh, N.~Schneider, N.~Hasabnis, T.~Mattson, Y.~Pinter, and G.~Oren, ``Advising openmp parallelization via a graph-based approach with transformers,'' \emph{arXiv preprint arXiv:2305.11999}, 2023.

\bibitem{schneider2023mpi}
\BIBentryALTinterwordspacing
N.~Schneider, T.~Kadosh, N.~Hasabnis, T.~Mattson, Y.~Pinter, and G.~Oren, ``{MPI-RICAL: Data-Driven MPI Distributed Parallelism Assistance with Transformers},'' in \emph{Proceedings of the SC '23 Workshops of The International Conference on High Performance Computing, Network, Storage, and Analysis}, ser. SC-W '23.\hskip 1em plus 0.5em minus 0.4em\relax New York, NY, USA: Association for Computing Machinery, 2023, p. 2–10. [Online]. Available: \url{https://doi.org/10.1145/3624062.3624063}
\BIBentrySTDinterwordspacing

\bibitem{schneider2024mpirigen}
N.~Schneider, N.~Hasabnis, V.~A. Vo, T.~Kadosh, N.~Krien, M.~Capota, G.~Tamir, T.~L. Willke, N.~Ahmed, Y.~Pinter \emph{et~al.}, ``Mpirigen: Mpi code generation through domain-specific language models,'' in \emph{Proceedings of the 2024 Workshop on AI For Systems}, 2024, pp. 1--6.

\bibitem{harel2023learning}
R.~Harel, Y.~Pinter, and G.~Oren, ``{Learning to parallelize in a shared-memory environment with transformers},'' in \emph{Proceedings of the 28th ACM SIGPLAN Annual Symposium on Principles and Practice of Parallel Programming}, 2023, pp. 450--452.

\bibitem{kadosh2024monocoder}
T.~Kadosh, N.~Hasabnis, V.~A. Vo, N.~Schneider, N.~Krien, M.~Capot{\u{a}}, A.~Wasay, G.~Tamir, T.~Willke, N.~Ahmed \emph{et~al.}, ``Monocoder: Domain-specific code language model for hpc codes and tasks,'' in \emph{2024 IEEE High Performance Extreme Computing Conference (HPEC)}.\hskip 1em plus 0.5em minus 0.4em\relax IEEE, 2024, pp. 1--7.

\bibitem{chen2024ompgpt}
L.~Chen, A.~Bhattacharjee, N.~Ahmed, N.~Hasabnis, G.~Oren, V.~Vo, and A.~Jannesari, ``{OMPGPT: A generative pre-trained transformer model for openmp},'' \emph{arXiv preprint arXiv:2401.16445}, 2024.

\bibitem{roziere2020unsupervised}
B.~Roziere, M.-A. Lachaux, L.~Chanussot, and G.~Lample, ``Unsupervised translation of programming languages,'' \emph{Advances in neural information processing systems}, vol.~33, pp. 20\,601--20\,611, 2020.

\bibitem{dhruv2025leveraging}
A.~Dhruv and A.~Dubey, ``Leveraging large language models for code translation and software development in scientific computing,'' in \emph{Proceedings of the Platform for Advanced Scientific Computing Conference}, 2025, pp. 1--9.

\bibitem{chen2024landscape}
L.~Chen, N.~K. Ahmed, A.~Dutta, A.~Bhattacharjee, S.~Yu, Q.~I. Mahmud, W.~Abebe, H.~Phan, A.~Sarkar, B.~Butler \emph{et~al.}, ``The landscape and challenges of hpc research and llms,'' \emph{arXiv preprint arXiv:2402.02018}, 2024.

\bibitem{huang2025paracoder}
X.~Huang, X.~Zhang, L.~Tao, R.~Mao, N.~Zhou, W.~Zhu, M.~Deng, J.~Meng, Y.~Wei, A.~C. Zhou \emph{et~al.}, ``Paracoder: Parallel code generation with large language model,'' in \emph{Proceedings of the 1st FastCode Programming Challenge}, 2025, pp. 1--7.

\bibitem{yadav5114036evaluating}
D.~Yadav and S.~Mondal, ``Evaluating pre-trained large language models on zero shot prompts for parallelization of source code,'' \emph{Available at SSRN 5114036}, 2025.

\bibitem{jain2025assessing}
R.~Jain and R.~Purandare, ``Assessing large language models in comprehending and verifying concurrent programs across memory models,'' \emph{arXiv preprint arXiv:2501.14326}, 2025.

\bibitem{yin2025chathpc}
J.~Yin, J.~Hines, E.~Herron, T.~Ghosal, H.~Liu, S.~Prentice, V.~Lama, and F.~Wang, ``chathpc: Empowering hpc users with large language models,'' \emph{The Journal of Supercomputing}, vol.~81, no.~1, p. 194, 2025.

\bibitem{mahmud2025autoparllm}
Q.~I. Mahmud, A.~TehraniJamsaz, H.~D. Phan, L.~Chen, M.~Capot{\u{a}}, T.~L. Willke, N.~K. Ahmed, and A.~Jannesari, ``Autoparllm: Gnn-guided context generation for zero-shot code parallelization using llms,'' in \emph{Proceedings of the 2025 Conference of the Nations of the Americas Chapter of the Association for Computational Linguistics: Human Language Technologies (Volume 1: Long Papers)}, 2025, pp. 11\,821--11\,841.

\bibitem{rahman2025performance}
A.~Rahman, V.~Cvetkovic, K.~Reece, A.~Walters, Y.~Hassan, A.~Tummeti, B.~Torres, D.~Cooney, M.~Ellis, and D.~Nikolopoulos, ``Performance evaluation of large language models for high-performance code generation: A multi-agent approach (marco),'' \emph{arXiv preprint arXiv:2505.03906v1}, 2025.

\bibitem{bolet2025can}
G.~Bolet, G.~Georgakoudis, H.~Menon, K.~Parasyris, N.~Hasabnis, H.~Estes, K.~W. Cameron, and G.~Oren, ``Can large language models predict parallel code performance?'' \emph{arXiv preprint arXiv:2505.03988}, 2025.

\bibitem{harel2025pragformer}
R.~Harel, T.~Kadosh, N.~Hasabnis, T.~Mattson, Y.~Pinter, and G.~Oren, ``Pragformer: data-driven parallel source code classification with transformers,'' \emph{International Journal of Parallel Programming}, vol.~53, no.~1, pp. 1--26, 2025.

\bibitem{nichols2024can}
D.~Nichols, J.~H. Davis, Z.~Xie, A.~Rajaram, and A.~Bhatele, ``Can large language models write parallel code?'' in \emph{Proceedings of the 33rd International Symposium on High-Performance Parallel and Distributed Computing}, 2024, pp. 281--294.

\bibitem{nichols2024hpc}
D.~Nichols, A.~Marathe, H.~Menon, T.~Gamblin, and A.~Bhatele, ``Hpc-coder: Modeling parallel programs using large language models,'' in \emph{ISC High Performance 2024 Research Paper Proceedings (39th International Conference)}.\hskip 1em plus 0.5em minus 0.4em\relax Prometeus GmbH, 2024, pp. 1--12.

\bibitem{chaturvedi2025hpc}
A.~Chaturvedi, D.~Nichols, S.~Singh, and A.~Bhatele, ``Hpc-coder-v2: Studying code llms across low-resource parallel languages,'' in \emph{ISC High Performance 2025 Research Paper Proceedings (40th International Conference)}.\hskip 1em plus 0.5em minus 0.4em\relax Prometeus GmbH, 2025, pp. 1--14.

\bibitem{nichols2025learning}
D.~Nichols, ``On learning behaviors of parallel code and systems across modalities,'' Ph.D. dissertation, University of Maryland, College Park, 2025.

\bibitem{jin2023benchmark}
Z.~Jin and J.~S. Vetter, ``A benchmark suite for improving performance portability of the sycl programming model,'' in \emph{2023 IEEE International Symposium on Performance Analysis of Systems and Software (ISPASS)}.\hskip 1em plus 0.5em minus 0.4em\relax IEEE, 2023, pp. 325--327.

\bibitem{liu2023lost}
N.~F. Liu, K.~Lin, J.~Hewitt, A.~Paranjape, M.~Bevilacqua, F.~Petroni, and P.~Liang, ``Lost in the middle: How language models use long contexts,'' \emph{arXiv preprint arXiv:2307.03172}, 2023.

\bibitem{kadosh2024ompar}
T.~Kadosh, N.~Hasabnis, P.~Soundararajan, V.~A. Vo, M.~Capota, N.~Ahmed, Y.~Pinter, and G.~Oren, ``Ompar: Automatic parallelization with ai-driven source-to-source compilation,'' \emph{arXiv preprint arXiv:2409.14771}, 2024.

\bibitem{yan2023codescope}
W.~Yan, H.~Liu, Y.~Wang, Y.~Li, Q.~Chen, W.~Wang, T.~Lin, W.~Zhao, L.~Zhu, H.~Sundaram \emph{et~al.}, ``Codescope: An execution-based multilingual multitask multidimensional benchmark for evaluating llms on code understanding and generation,'' \emph{arXiv preprint arXiv:2311.08588}, 2023.

\bibitem{ashrafi2025enhancing}
N.~Ashrafi, S.~Bouktif, and M.~Mediani, ``Enhancing llm code generation: A systematic evaluation of multi-agent collaboration and runtime debugging for improved accuracy, reliability, and latency,'' \emph{arXiv preprint arXiv:2505.02133}, 2025.

\bibitem{achiam2023gpt}
J.~Achiam, S.~Adler, S.~Agarwal, L.~Ahmad, I.~Akkaya, F.~L. Aleman, D.~Almeida, J.~Altenschmidt, S.~Altman, S.~Anadkat \emph{et~al.}, ``Gpt-4 technical report,'' \emph{arXiv preprint arXiv:2303.08774}, 2023.

\bibitem{grattafiori2024llama}
A.~Grattafiori, A.~Dubey, A.~Jauhri, A.~Pandey, A.~Kadian, A.~Al-Dahle, A.~Letman, A.~Mathur, A.~Schelten, A.~Vaughan \emph{et~al.}, ``The llama 3 herd of models,'' \emph{arXiv preprint arXiv:2407.21783}, 2024.

\bibitem{chen2021evaluating}
M.~Chen, J.~Tworek, H.~Jun, Q.~Yuan, H.~P. d.~O. Pinto, J.~Kaplan, H.~Edwards, Y.~Burda, N.~Joseph, G.~Brockman \emph{et~al.}, ``Evaluating large language models trained on code,'' \emph{arXiv preprint arXiv:2107.03374}, 2021.

\bibitem{hendrycks2020measuring}
D.~Hendrycks, C.~Burns, S.~Basart, A.~Zou, M.~Mazeika, D.~Song, and J.~Steinhardt, ``Measuring massive multitask language understanding,'' \emph{arXiv preprint arXiv:2009.03300}, 2020.

\bibitem{micikevicius2017mixed}
P.~Micikevicius, S.~Narang, J.~Alben, G.~Diamos, E.~Elsen, D.~Garcia, B.~Ginsburg, M.~Houston, O.~Kuchaiev, G.~Venkatesh \emph{et~al.}, ``Mixed precision training,'' \emph{arXiv preprint arXiv:1710.03740}, 2017.

\bibitem{dettmers2023qlora}
T.~Dettmers, A.~Pagnoni, A.~Holtzman, and L.~Zettlemoyer, ``Qlora: Efficient finetuning of quantized llms,'' \emph{Advances in neural information processing systems}, vol.~36, pp. 10\,088--10\,115, 2023.

\bibitem{azureopenai}
{Microsoft Corporation}, ``{Azure OpenAI Service},'' \url{https://azure.microsoft.com/en-us/products/ai-services/openai-service}, accessed: 2024-06-30.

\bibitem{austin2021program}
J.~Austin, A.~Odena, M.~Nye, M.~Bosma, H.~Michalewski, D.~Dohan, E.~Jiang, C.~Cai, M.~Terry, Q.~Le \emph{et~al.}, ``Program synthesis with large language models,'' \emph{arXiv preprint arXiv:2108.07732}, 2021.

\bibitem{radford2019language}
A.~Radford, J.~Wu, R.~Child, D.~Luan, D.~Amodei, I.~Sutskever \emph{et~al.}, ``Language models are unsupervised multitask learners,'' \emph{OpenAI blog}, vol.~1, no.~8, p.~9, 2019.

\bibitem{bareiss2022code}
P.~Barei{\ss}, B.~Souza, M.~d'Amorim, and M.~Pradel, ``Code generation tools (almost) for free? a study of few-shot, pre-trained language models on code,'' \emph{arXiv preprint arXiv:2206.01335}, 2022.

\bibitem{xu2024does}
D.~Xu, T.~Xie, B.~Xia, H.~Li, Y.~Bai, Y.~Sun, and W.~Wang, ``Does few-shot learning help llm performance in code synthesis?'' \emph{arXiv preprint arXiv:2412.02906}, 2024.

\bibitem{kadosh2023monocoder}
T.~Kadosh, N.~Hasabnis, V.~A. Vo, N.~Schneider, N.~Krien, M.~Capota, A.~Wasay, N.~Ahmed, T.~Willke, G.~Tamir \emph{et~al.}, ``Monocoder: Domain-specific code language model for hpc codes and tasks,'' \emph{arXiv preprint arXiv:2312.13322}, 2023.

\bibitem{chaturvedi2024hpc}
A.~Chaturvedi, D.~Nichols, S.~Singh, and A.~Bhatele, ``Hpc-coder-v2: Studying code llms across low-resource parallel languages,'' \emph{arXiv preprint arXiv:2412.15178}, 2024.

\bibitem{shengyu2023instruction}
Z.~Shengyu, D.~Linfeng, L.~Xiaoya, Z.~Sen, S.~Xiaofei, W.~Shuhe, L.~Jiwei, R.~Hu, Z.~Tianwei, F.~Wu \emph{et~al.}, ``Instruction tuning for large language models: A survey,'' \emph{arXiv preprint arXiv:2308.10792}, 2023.

\bibitem{wu2025shadow}
T.~Wu, R.~Yang, J.~Li, P.~Hu, N.~Wong, and Y.~Yang, ``Shadow-ft: Tuning instruct via base,'' \emph{arXiv preprint arXiv:2505.12716}, 2025.

\bibitem{torchtune}
\BIBentryALTinterwordspacing
torchtune maintainers and contributors, ``torchtune: Pytorch's finetuning library,'' Apr. 2024. [Online]. Available: \url{https//github.com/pytorch/torchtune}
\BIBentrySTDinterwordspacing

\bibitem{chen2023compcodevet}
L.~Chen, A.~Bhattacharjee, N.~K. Ahmed, N.~Hasabnis, G.~Oren, B.~Lei, and A.~Jannesari, ``{Compcodevet: A compiler-guided validation and enhancement approach for code dataset},'' \emph{arXiv preprint arXiv:2311.06505}, 2023.

\bibitem{wang2022compilable}
X.~Wang, Y.~Wang, Y.~Wan, F.~Mi, Y.~Li, P.~Zhou, J.~Liu, H.~Wu, X.~Jiang, and Q.~Liu, ``Compilable neural code generation with compiler feedback,'' \emph{arXiv preprint arXiv:2203.05132}, 2022.

\bibitem{jana2024cotran}
P.~Jana, P.~Jha, H.~Ju, G.~Kishore, A.~Mahajan, and V.~Ganesh, ``Cotran: An llm-based code translator using reinforcement learning with feedback from compiler and symbolic execution,'' in \emph{ECAI 2024}.\hskip 1em plus 0.5em minus 0.4em\relax IOS Press, 2024, pp. 4011--4018.

\bibitem{chen2024fortran2cpp}
L.~Chen, B.~Lei, D.~Zhou, P.-H. Lin, C.~Liao, C.~Ding, and A.~Jannesari, ``Fortran2cpp: Automating fortran-to-c++ migration using llms via multi-turn dialogue and dual-agent integration,'' \emph{arXiv preprint arXiv:2412.19770}, 2024.

\bibitem{dearing2024lassi}
M.~T. Dearing, Y.~Tao, X.~Wu, Z.~Lan, and V.~Taylor, ``{LASSI}: An llm-based automated self-correcting pipeline for translating parallel scientific codes,'' \emph{arXiv preprint arXiv:2407.01638}, 2024.

\end{thebibliography}
